\documentclass[manuscript]{acmart}

\usepackage{booktabs}
\usepackage{longtable}
\usepackage{array}
\usepackage{tabularx}
\usepackage{multirow}
\usepackage{graphicx}
\usepackage{xurl}
\usepackage{placeins}
\title[Spatial Dissonance in AR Tour Reception]{“The Rose Garden Is a Cemetery”: Spatial Dissonance and Divergent Public Experiences of an Indigenous Augmented Reality Tour}
\author{Grace Fujii}
\author{Vittorio Murtas}
\author{Sarah Sallee}
\author{Isabella 'Amne Gomez}
\author{Lauren Baines}
\author{Danielle Heitmuller}
\author{Amy Lueck}
\author{Kai Lukoff}

\setcopyright{none}
\acmYear{2026}
\copyrightyear{2026}
\acmDOI{}
\acmISBN{}
\renewcommand\footnotetextcopyrightpermission[1]{}


\begin{abstract}
Community-authored augmented reality (AR) can contest dominant heritage narratives by juxtaposing stories with physical surroundings. We examine the Thámien Ohlone AR Tour, developed with members of the Ohlone Native American community at Mission Santa Clara and Santa Clara University. Combining paired pre/post surveys from 57 participants with collaborative thematic analysis, we compare reception among Ohlone community members, university affiliates, and Mission visitors. Spatial dissonance captures felt tensions when AR challenged familiar campus readings or when surroundings conflicted with Ohlone histories the AR affirmed, such as a garden's significance as a burial ground. Cultural frames shaped interpretation; contemporary people and relationships made Ohlone continuance tangible. Participants appreciated the tour and called for permanent physical acknowledgment. We propose staging reflective encounters, clarifying community-authored meanings, and using AR to prototype public acknowledgment and support community-directed advocacy. These findings foreground relationships to land and place, informed by Indigenous scholarship, in designing and evaluating AR.

\end{abstract}

\begin{document}
\maketitle
\section{Introduction}\label{sec:1}

Beside the Mission Santa Clara church, at the center of the Santa Clara University (SCU) campus, a low wall and a locked gate enclose what looks like a rose garden. It is a cemetery, one of several burial grounds on and beside the campus that together hold more than 7,000 Native Californians \citep{ohwg2020report}. More than a century of archaeology and commemoration has made the campus read as a settler heritage landmark while keeping Indigenous presence out of view \citep{panich2022archaeology}. In 2020, a university working group that included Ohlone leaders recommended plaques and interpretive panels at all campus burial sites \citep{ohwg2020report}. Subsequent initiatives included removing contested monuments and creating a permanent digital memorial in the campus museum \citep{nasr2020rallying, calande2023those}. Yet permanent, on-site interpretation remained limited: during our study, no signage at the cemetery entrance identified it as a Native burial ground or as Ohlone land.

Augmented reality (AR) offers a way to contest whose histories are represented in public space. Artists and activists have long used temporary media to change what a monument says without touching it: projection artists cast suppressed histories onto the facades of monuments \citep{wodiczko1999critical}, and guerrilla AR carried the move into the digital layer, planting unsanctioned messages on logos, galleries, and public squares \citep{skwarek2018activism}. AR activists now erect geofenced digital monuments and reveal invisible histories in place \citep{silva2022understanding}, retell the story of a site "with or without permission of its owner" \citep{silva2024permission}, and render on the ground the plaques a community wants to see there \citep{schroeder2023location}. For a landscape like this campus, the AR medium can represent the ground as the community understands it now, before physical changes are made.

The Thámien Ohlone AR Tour belongs to this practice. Created in collaboration with Ohlone Native American community members by students and faculty at SCU, and supported by the Muwekma Ohlone Tribe \citep{jauregui2024wearestillhere, lukoff2025community}, the version evaluated in this study anchors narration by Native storytellers and 3D reconstructions to seven campus stops, from life-size tule houses at the unmarked village site to the names of mission-era Ohlone individuals displayed in the Chochenyo language above the church. Its content carries five messages the community defined, among them "This is Ohlone Land" and "We are Still Here": an exercise of rhetorical sovereignty, the right of Native peoples to determine their own communicative goals and modes \citep{lyons2000rhetorical}, extended into digital form \citep{haas2005making}.

We compared three overlapping publics because the site is simultaneously Ohlone ancestral land, a university campus, and a Catholic mission. Ohlone community members could bring ancestral relationships and concerns about cultural representation; university affiliates, everyday familiarity and institutional ties; and Mission visitors, religious commitments or interests in mission heritage. These relationships offered different starting points for encountering Ohlone accounts of cultural life, colonial harm, and continuance. Following reception theory \citep{hall1980encoding}, we examined how participants drew on these resources to interpret the same community-authored representations, without assuming that public membership determined their views.

Understanding what community-authored AR contributes at contested sites requires examining how people receive its messages. Drawing on interviews with AR creators, \citet{silva2022understanding} identify audience responses as a direction for future research. We examine those responses across publics with different relationships to the same site, asking how participants interpret community-defined messages in the places those messages concern. This comparison attends to shared responses as well as differences in interpretation, emotion, and expectations for the site's future.

We conducted a field study and survey of the Thámien Ohlone AR Tour with all three of its publics: Ohlone community members (n=11), university community members (n=24), and Mission visitors (n=25), combining pre/post surveys with in-situ interviews and observations, analyzed through collaborative thematic analysis. Four Native participants who are not Ohlone also took part; we report their responses separately rather than counting them as Ohlone. We asked:

\begin{itemize}
\item
  \textbf{RQ1.} How do different communities receive and engage with the tour\textquotesingle s community-defined messages?
\item
  \textbf{RQ2.} What factors shape differences in message reception across communities?
\item
  \textbf{RQ3.} What design implications emerge from different publics' reception of community-authored AR at culturally contested sites?
\end{itemize}

The tour's reception shows how relationships to land and place shape what community-authored AR makes visible and what participants ask of it. At the cemetery, a serene garden and an unmarked burial ground remained present together, prompting shock, grief, and questions about the campus's treatment of Ohlone history. We describe this felt tension between the campus's visible presentation and its meanings as spatial dissonance. For university participants and some Mission visitors, AR challenged everyday or Catholic readings of the campus. For Ohlone participants, it could affirm ancestral connections while making the campus's limited acknowledgment of Ohlone history more troubling. Cultural frames also shaped interpretation: familiar stories could help participants engage with the Ohlone account or absorb it into another narrative, while contemporary people and family relationships made continued Ohlone presence tangible. Across the three publics, appreciation of the tour accompanied calls for physical acknowledgment, connecting its representations to questions about what the campus should become.

This paper contributes a comparative account of how overlapping publics receive community-authored AR at a contested site, and uses spatial dissonance to examine the emotional force of tensions between a site's visible presentation and its meanings. We translate this account into three design directions: staging encounters that allow time for observation, reveal, and reflection; answering ``why here?'' by making community-authored meanings and their ties to place and contemporary life legible; and using AR to prototype public acknowledgment and support community-directed advocacy. Together, these directions connect immediate reception with possibilities for the site's future under community direction.

\section{The Thámien Ohlone Augmented Reality Tour}\label{sec:2}

To situate the reception study, we describe the tour experience and the five messages Ohlone collaborators defined for visitors.

\begin{figure}[htbp]
\centering
\includegraphics[width=\linewidth]{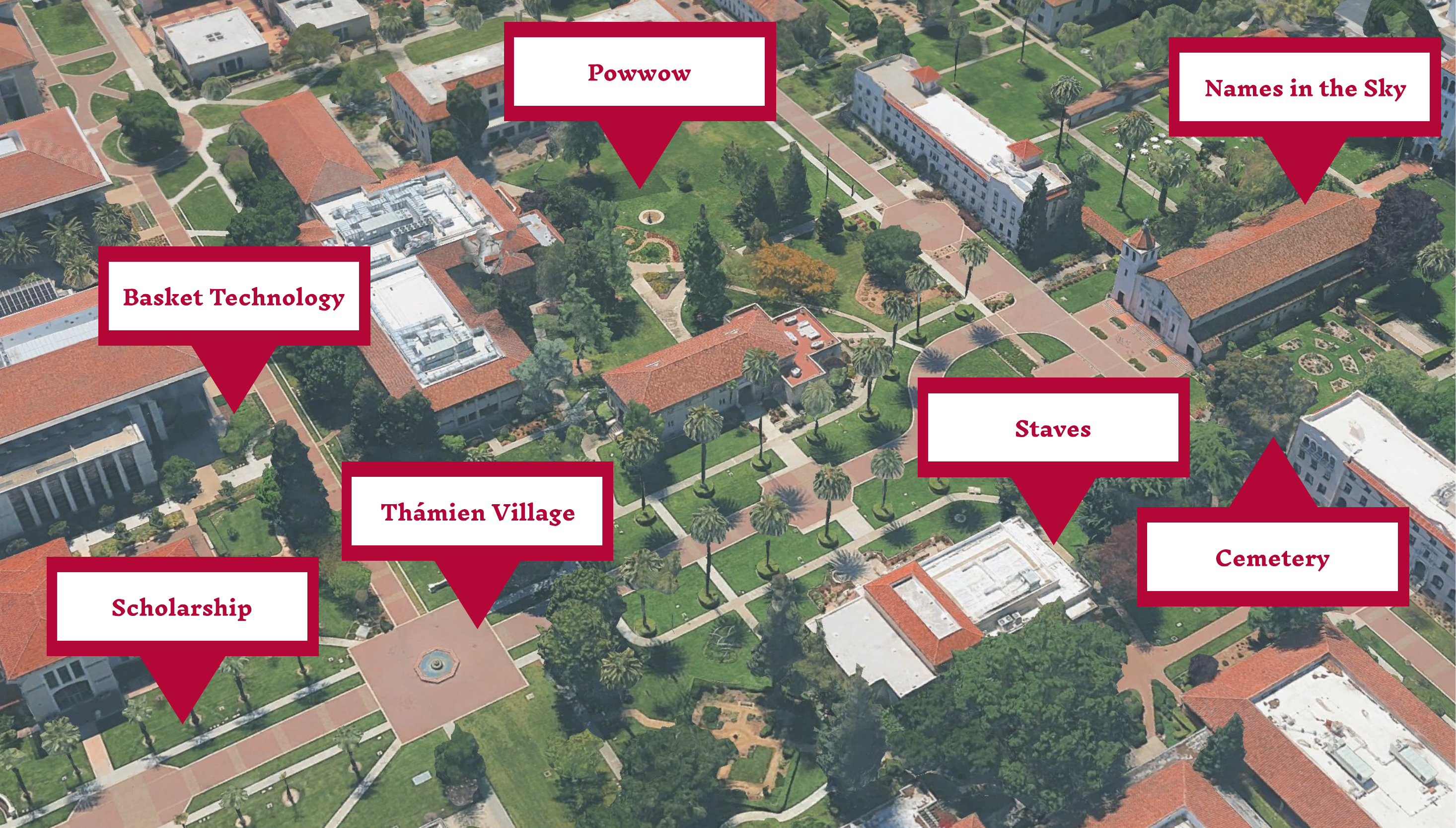}
\caption{The seven stops of the Thámien Ohlone AR Tour on the Santa Clara University campus during the 2026 study.}
\label{fig:tour-map}
\Description{An oblique aerial campus map marks seven stops with labeled red pointers. Scholarship and Basket Technology appear on the left, Thámien Village near the lower center, Powwow above it, Staves to the right of center, and Names in the Sky and Cemetery on the right near the mission church. The stops are distributed among campus walkways, lawns, buildings, and the garden; no route connecting them is drawn.}
\end{figure}

\subsection{The tour}\label{sec:2-1}

The Thámien Ohlone AR Tour is a location-based augmented reality walking tour of the university campus, delivered as an app for iOS and Android. The campus stands on the site of Thámien, an Ohlone village inhabited for more than two thousand years before Mission Santa Clara de Asís was founded there in 1777. On opening the app, a visitor hears a welcome from the Muwekma Ohlone Tribe in Chochenyo and English (``We Ohlone have been here for generations and we are still here today'') and sees a map of the stops. At each stop the visitor points the phone at a landmark so that the app can localize, and the app plays narration recorded by an Ohlone community member while it overlays 3D models, photographs, and sound on the camera view. The version evaluated in this study was developed in Unity and used Niantic's Lightship Visual Positioning System (VPS) for localization. VPS matched camera imagery to previously mapped features of each site, enabling the app to align virtual content with the physical surroundings. Each stop is tied to its location, tells a single story, and has captions. Most stops take about two to three minutes to experience. The first three stops launched in October 2023; by the time of our study in 2026 the app held seven, released one or two at a time at community events (Table~\ref{tab:tour-stops}). A short video demonstrates the first three stops \citep{jauregui2024wearestillhere}. Position papers discuss social and technical considerations for choosing anchors in an earlier WebAR version of the tour \citep{nguyen2024sociotechnical} and the tour's co-design and community empowerment goals \citep{lukoff2025community}.

\begingroup
\small
\setlength{\tabcolsep}{4pt}
\renewcommand{\arraystretch}{1.12}
\begin{longtable}{@{}>{\raggedright\arraybackslash}p{\dimexpr.17\linewidth-1.5\tabcolsep\relax}>{\raggedright\arraybackslash}p{\dimexpr.25\linewidth-1.5\tabcolsep\relax}>{\raggedright\arraybackslash}p{\dimexpr.42\linewidth-1.5\tabcolsep\relax}>{\raggedright\arraybackslash}p{\dimexpr.16\linewidth-1.5\tabcolsep\relax}@{}}
\caption{The seven stops as of the 2026 study, in release order.}\label{tab:tour-stops}\\
\toprule
\textbf{Stop (released)} & \textbf{Site} & \textbf{What the visitor sees and hears} & \textbf{Message} \\
\midrule
\endfirsthead
\multicolumn{4}{@{}l}{\small\tablename~\thetable\ (continued)}\\
\toprule
\textbf{Stop (released)} & \textbf{Site} & \textbf{What the visitor sees and hears} & \textbf{Message} \\
\midrule
\endhead
\midrule
\multicolumn{4}{r@{}}{\small Continued on next page}\\
\endfoot
\bottomrule
\endlastfoot
Thámien Village Stop (Oct 2023) & Central fountain at Thámien's archaeologically identified center; unmarked. & Life-size tule houses visitors can enter, a village illustration, and a soundscape of daily life. & This is Ohlone land \\
\addlinespace
Names in the Sky Stop (Oct 2023) & Mission Santa Clara church facade. & The known names of 2,475 ancestors buried at the mission rise in Chochenyo as narration recounts mission life, death, and Muwekma Ohlone survival. & Recovering from colonialism \\
\addlinespace
Powwow Stop (Oct 2023) & Lawn hosting the annual campus powwow, an intertribal gathering featuring music and dance. & Past powwow photographs and a Chochenyo healing song recorded by Jose Guzman in the 1930s and five tribal members in 2023; his great-granddaughter narrates. & We are still here \\
\addlinespace
Cemetery Stop (May 2024) & Mission Rose Garden beside the church; about 2,400 unmarked Native burials. & Red abalone shells appear across the garden. Narration explains their meaning in Ohlone burial tradition and the site's Chochenyo renaming by the Muwekma Ohlone Tribe's Language Committee. & Recovering from colonialism \\
\addlinespace
Basket Technology Stop (Oct 2024) & Benches beneath a willow beside the engineering building. & Visitors tap virtual tule, willow, and olivella shell beads to learn their uses and view a 3D-scanned basket. An Ohlone weaver who calls herself a ``traditional technologist'' narrates. & Ohlone culture is dynamic \\
\addlinespace
Staves Stop (May 2025) & Museum lawn used for the Ohlone youth cultural camp. & Visitors play Uttkaani Tapporikma on the team of one of two Muwekma Ohlone youth ambassadors, who are cousins, using 3D scans of their grandfather's hand-carved set. & Ohlone culture is dynamic \\
\addlinespace
Scholarship Stop (May 2026) & Admissions building entrance. & Seedlings grow into trees meaningful to two Native student recipients of the Ohlone and Muwekma Ohlone Scholarship, who describe studying on this land. & We are still here \\
\end{longtable}
\endgroup

\subsection{Co-design and the five messages}\label{sec:2-2}

The tour was developed by SCU students and faculty (the design team) with Ohlone community members, including members of the Muwekma Ohlone Tribe, which has supported the project throughout. It builds on longstanding Ohlone collaborations with the campus museum and university researchers \citep{panich2024fighting, lueck2021public}. Published accounts describe Ohlone partners as shaping the tour's stories and its emphasis on contemporary Ohlone presence \citep{lukoff2025community, calande2023revealing}. The team's project documentation, shared with us, lists five messages that the Ohlone partners defined for visitors:

\begin{itemize}
\item
  \emph{This is Ohlone land:} an alternative history of Santa Clara grounded in Ohlone expertise and the campus's historical and archaeological record.
\item
  \emph{Recovering from colonialism:} ongoing colonial harm and Native and non-Native efforts to address it through historical research and revitalization.
\item
  \emph{Ohlone culture is dynamic:} cultural change does not imply a loss of authenticity.
\item
  \emph{Positionality:} differences in Ohlone experience shaped by age, gender, status, and their intersections.
\item
  \emph{We are still here:} concrete ways to support Ohlone and Muwekma Ohlone culture and continued presence on their homelands.
\end{itemize}

Published accounts describe a process guided by design justice \citep{costanzachock2020design}, the disability-rights principle ``nothing about us without us'' \citep{charlton1998nothing}, and rhetorical sovereignty, Native peoples' right to tell their own stories for their own purposes \citep{lyons2000rhetorical}. Ohlone partners, faculty, and students jointly proposed and prioritized stops on a campus map \citep{jauregui2024wearestillhere}, asking ``Why here?'' when selecting locations \citep{nguyen2024sociotechnical}. Tour-stop prototypes were iteratively tested and revised with feedback from Ohlone co-designers and other community members. The design team also held biweekly AR skills workshops with Ohlone youth and other community members \citep{lukoff2025community}.

\section{Related Work}\label{sec:3}

We draw on research on contested heritage, place and Indigenous relationships to land, audience reception, and Indigenous HCI. These perspectives guide our analysis of how people with different ties to the same site interpret community-authored AR.

\subsection{Contested heritage sites and their publics}\label{sec:3-1}

Heritage studies begin from the observation that commemoration serves some claimants and disinherits others. \citet[p.~21]{tunbridge1996dissonant} treat the resulting dissonance as intrinsic: "all heritage is someone\textquotesingle s heritage and therefore logically not someone else\textquotesingle s," so dissonance is "a condition, whether active or latent, of all heritage." An authorized heritage discourse decides which pasts count \citep{smith2006uses}. Across interviews at 45 museums and heritage sites, \citet[pp.~2--6]{smith2021emotional} found reinforcement of existing feelings and commitments to be a dominant pattern. At the former Nazi Party Rally Grounds in Nuremberg, residents interviewed in the 2000s used the site for tennis and recreation; as one put it, "You get used to it. And then you don\textquotesingle t think about it any more" \citep[p.~181]{macdonald2009difficult}. At plantation museums, docents script historical empathy toward the planter class \citep{modlin2011tour}, and at one South Carolina plantation some visitors accepted the site\textquotesingle s celebratory account of planter life while others rejected it outright, with the split tracking the assumptions each visitor shared with the communities they came from \citep{buzinde2009interpreting}. Campus counter-tours make the university itself the contested site. Green describes leading the Hallowed Grounds tour of the University of Alabama's slave-built landscape as ``physically and emotionally taxing,'' and presents virtual tours as ``a sustainable alternative'' while calling for institutional funding and paid guides \citep[pp.~31, 40]{green2020burden}.

Mission Santa Clara brings these tensions together: a Catholic church and university landmark stands on Ohlone ancestral land beside Native burial grounds. Archaeology and commemoration have helped sustain a settler account of the campus while obscuring Indigenous presence \citep{panich2022archaeology}. A 2019 survey of 215 SCU undergraduates found low self-reported knowledge of campus heritage, including Indigenous histories, alongside interest in learning more \citep[pp.~139--142]{kroot2020students}. Community-engaged coursework has used digital media to challenge existing representations \citep{lueck2020representing, lueck2021public}. Physical acknowledgment was also a community priority: in 2020, a working group including Ohlone leaders recommended plaques and interpretive panels at all campus burial sites \citep{ohwg2020report}. During our study, the entrance still lacked signage identifying the cemetery. This work motivates attending to how ancestral connections, everyday campus life, and religious commitments shape encounters with the same place.

\subsection{Indigenous storytelling, land, and place}\label{sec:3-2}

Indigenous storytelling connects knowledge with land and community relationships. \citeauthor{basso1996wisdom}'s \citep{basso1996wisdom} study of Western Apache place-names shows how named places evoke histories and moral teachings. \citeauthor{archibald2008indigenous}'s \citep{archibald2008indigenous} Indigenous storywork, developed with Elders and storytellers, situates learning through stories within principles including respect, responsibility, and reciprocity. \citet{simpson2014land} describes Nishnaabeg learning through relationships with land; in \emph{Braiding Sweetgrass}, \citet[p.~25]{kimmerer2013braiding} presents plants as teachers whose gifts call for reciprocal care. AR projects also bring Indigenous place-names into view: iSikte Sápmi, developed by a Sámi language centre, reveals Sámi place-names in the surrounding landscape through a phone camera \citep{cocq2017traditional}, while OurWorlds, founded by a member of the Pala Band of Mission Indians, overlays Indigenous territory names on the ground a user views \citep{ourworlds2025app, mapp2022app}. Places and their names can recall stories, while the living land can itself offer lessons in how to relate to and care for it. For AR, this suggests examining how visitors connect a story to the ground before them, including what they understand about its significance to the community and how it should be treated.

\citet{harrison1996replacing} distinguish physical space from the social meanings and practices that make it a place. A house provides shelter, but becomes a home through the meanings and routines of the people who live there. AR designers have worked with place-meaning in two ways. The first builds on the meaning a visitor already holds. An early proposal for AR at heritage sites borrowed Walter Benjamin\textquotesingle s word aura for "the cultural and personal significance that a place (or object) holds for an individual," and argued that a mixed-reality experience "can exploit" that significance and add to it \citep[pp.~36, 39]{macintyre2004presence}. The second disturbs the meaning on purpose. A close precedent is \emph{Ghosts of the Horseshoe}, a campus AR project addressing the history of enslavement at the University of South Carolina. \citet[p.~33]{walker2014ghosts} describes content loading upon arrival with the intention to ``\,`shock' or surprise participants into awareness.'' \citet[p.~144]{levine2014common} defines spatial dissonance as experiencing two contradictory spaces or ideas about a space. \citet{jones2017augmented} apply this concept to AR ghost-bike memorials designed to unsettle routine perceptions of roadways. Design research has treated making the familiar strange \citep{bell2005making}, leaving meaning open for the user to complete \citep{gaver2003ambiguity}, and staging discomfort deliberately and briefly \citep{benford2012uncomfortable} as resources in their own right. One location-based AR experience invited people to lie alongside a creek and view its pre-colonial, industrial, and speculative future states; fifteen participants moved between human-centered and creek-centered readings of the same water \citep{yang2025beingcreek}. We examine what participants find surprising or unsettling when AR or the physical site conflicts with their understanding of place, and how these encounters differ across publics.

\subsection{Reception: what audiences bring, and who carries a message}\label{sec:3-3}

A message is not always read as it was intended. In \citeauthor{hall1980encoding}'s \citep{hall1980encoding} account of how media are received, a producer encodes a message with a preferred meaning, and each audience member decodes it from their own position: taking the preferred meaning as given (a dominant reading), accepting it in general while adjusting it to their own situation (a negotiated reading), or rejecting its frame altogether (an oppositional reading). Hall developed these positions in relation to hegemonic broadcasting. Here we adapt them to describe acceptance, negotiation, or resistance to the community-authored account: a visitor can resist the tour's message while drawing on the dominant settler account of the mission. A related tension appears in museum documentation. Comparing Cambridge museum catalog entries with Zuni community members' accounts of the same objects, \citet{srinivasan2010diverse} found that the catalog omitted stories and information about the objects' historical and contemporary uses. They proposed a digital system in which community-controlled accounts would have the same status as museum descriptions. \citet[pp.~138, 149, 156]{potter1987discourse} describe interpretive repertoires as recurrent ways of describing and evaluating events that people draw on flexibly across situations. We use this concept to examine how familiar ways of speaking can support different readings of the tour. A study at a science center found that visitors' learning varied with prior knowledge, motivation, and expectations \citep{falk2005contextual}.

We ask how audiences' prior knowledge and expectations shape their reading of the tour's message ``we are still here.'' This message connects to survivance: active Native presence and the continuation of stories that counter narratives of disappearance \citep{vizenor2008survivance}. Related studies examine how audiences encounter Indigenous histories: tribal college students discussed representation, land dispossession, and community resilience while playing \emph{When Rivers Were Trails} \citep{emmons2021ancestors}; non-Indigenous undergraduates who encountered residential-school Survivor testimony in text or VR reported greater empathy and political solidarity than a control group immediately afterward; VR did not significantly outperform text, and the effects declined over time \citep{starzyk2024influence}. We compare how publics with different relationships to the site interpret the same Ohlone accounts.

\subsection{Indigenous future imaginaries and AR}\label{sec:3-4}

HCI research with Indigenous communities offers approaches to designing technology around community priorities. Recent CHI papers describe an environmental data platform co-designed with Ojibwe partners to support treaty rights and the stewardship of manoomin, or wild rice \citep{mckenna2026noondawind}, and participatory work with Radio Tosepan Limakxtum toward Taewaloni, an envisioned Masewal community-led database grounded in relations with land as a living entity \citep{guerreromillan2026digital}. That strand also warns against designing for a generalized Indigenous community, since what counts as respectful practice differs from one community to the next \citep{noe2024generalized}. Related work situates design within the histories that shape it \citep{irani2010postcolonial} and documents how Native American advocates use Twitter to circulate political information and connect around shared issues \citep{vigilhayes2017indigenous}. Building on these approaches, we examine how a community's account of a place is received by people with different relationships to that place.

The tour studied here was built with Ohlone community members, who defined its five messages in co-design workshops, partake in design and development sessions, and narrate its stops. Its message "we are still here" asserts a continuing presence, a claim on the future as well as the present. That message voices a future imaginary, "a vision of the future that is shared by a group of people and used to motivate change in the present" \citep[p.~11]{lewis2023future}. Lewis connects these imaginaries to sovereignty through Scott Richard Lyons's formulation: ``The pursuit of sovereignty is to revise not our past, but our possibilities'' \citep[quoted in~][p.~18]{lewis2023future}.

AR activism, AR made for a social cause \citep{skwarek2018activism, silva2022understanding}, lets activists "re-tell the story of a place, with or without permission of its owner" \citep[p.~1]{silva2024permission}, including unpermitted "guerilla installations at museums exposing questionable art donations" \citep[p.~12]{silva2022understanding}. After Chanel Miller was sexually assaulted on Stanford\textquotesingle s campus, the university agreed to a contemplative garden at the site, but rejected the quotes from her that she proposed for its plaque, and left it "unmarked for several years" \citep[p.~2]{schroeder2023location}. The AR project Dear Visitor placed a virtual plaque there with one of those quotes, and its makers believe the physical plaque Stanford installed in 2020 "took direct inspiration" from it \citep[p.~3]{schroeder2023location}. On campuses, physical change is slow, costly, and subject to approval; projects that use ephemeral performance and digital space "sidestep the challenge of institutional inertia or resistance" \citep[p.~13]{lueck2021public}. At the Cemetery Stop, the tour\textquotesingle s message "this is Ohlone land" identified the ground in terms absent from the entrance signage during our study. A future imaginary, in Lewis\textquotesingle s definition, has to be shared. A claim that arrives this way, without waiting on a marker, therefore depends on the people who encounter it taking it up, which matters for design in three ways.

First, visitors bring their own interpretive frameworks and may negotiate or oppose a tour's intended meanings \citep{hall1980encoding}. Second, reception research can examine how Ohlone community members who did not help design the tour interpret the messages its Ohlone collaborators chose. Third, if a person\textquotesingle s relationship to the place also shapes what they take away, designers could tailor content by group, since the same content "may be just right for some, and totally miss the mark for others" \citep[p.~122]{falk2013understanding}. A study of twenty AR activism creators asks for exactly such research: future work "should consider interviewing or surveying audience members" \citep[p.~13]{silva2022understanding}. This paper does so with Ohlone community members, who can judge whether the messages came through; students and staff, who cross the campus daily and for whom the site is ordinary ground rather than a destination; and visitors who came to hear Mission Santa Clara\textquotesingle s own telling of its history, which the tour contests. The findings compare how the public encountered the site, interpreted the messages, and articulated expectations for what should follow.

\section{Methods}\label{sec:4}

Between April and July 2026, we conducted a mixed-methods field study of the Thámien Ohlone AR Tour at Mission Santa Clara and the surrounding university campus. We combined pre- and post-tour surveys with observation, audio-recorded conversations during the tour, and an Ohlone community debrief. The surveys describe participants' self-assessments and responses to the experience; the qualitative analysis examines how participants interpreted the community-defined messages and related them to the site. We analyzed these materials separately, then interpreted their patterns together.

\subsection{Recruitment and participants}\label{sec:4-1}

Fifty-seven participants completed the six paired survey items and four common post-tour items and formed the quantitative sample summarized in Table~\ref{tab:participants}. Four additional participants had incomplete post-tour surveys; three contributed qualitative material. Appendix~\ref{app:b1} accounts for all 61 participants retained in the survey sample.

We compared three overlapping publics: Ohlone community members, university community members, and Mission visitors. We use \emph{public} to describe a shared relationship to the site. Three participants belonged to both the Ohlone and university publics, and four to both the university and Mission-visitor publics. Four Native participants who did not identify as Ohlone are reported separately, preserving distinctions among Indigenous communities and their relationships to this land \citep{noe2024generalized}. Their small number limits comparative interpretation.

\begin{table}[htbp]
\caption{Participants with complete surveys, by group, age, and recruitment.}
\label{tab:participants}
\centering
\small
\setlength{\tabcolsep}{3.5pt}
\renewcommand{\arraystretch}{1.15}
\begin{tabularx}{\linewidth}{@{}>{\raggedright\arraybackslash}p{.25\linewidth}>{\centering\arraybackslash}p{.035\linewidth}>{\centering\arraybackslash}p{.07\linewidth}>{\centering\arraybackslash}p{.06\linewidth}>{\centering\arraybackslash}p{.05\linewidth}>{\raggedright\arraybackslash}X@{}}
\toprule
\multirow[t]{2}{=}{\textbf{Participant group}} & \multirow[t]{2}{=}{\textbf{n}} & \multicolumn{3}{c}{\textbf{Age (years), n}} & \multirow[t]{2}{=}{\textbf{Recruitment and composition}} \\
\cmidrule(lr){3-5}
& & \textbf{Under 18} & \textbf{18--24} & \textbf{25+} & \\
\midrule
Ohlone community members & 11 & 6 & 1 & 4 & Campus powwow and Ohlone Youth Cultural Camp. \\
\addlinespace
Native participants who were not Ohlone & 4 & 0 & 1 & 3 & Campus powwow. \\
\addlinespace
University community members & 24 & 1 & 17 & 5 & Primarily campus word of mouth; mostly undergraduate and graduate students, alongside staff, alumni, and parents accompanying students. \\
\addlinespace
Mission visitors & 25 & 9 & 2 & 14 & High-school pilgrimage group, Sunday Mass, mission staff, and local mission-related networks. \\
\midrule
\textbf{Unique participants} & \textbf{57} & \textbf{15} & \textbf{18} & \textbf{23} & --- \\
\bottomrule
\end{tabularx}
\par\smallskip
\begin{minipage}{\linewidth}
\footnotesize\textit{Note. Counts describe the 57 participants with complete surveys. Public memberships overlap, so group rows should not be summed. Age was unavailable for one university participant.}
\end{minipage}
\end{table}

Recruitment relied on convenience sampling and existing events. Participants varied in age, voluntariness, prior relationship to the site, and touring-group size; these differences inform interpretation of the comparisons across publics.

\subsection{Tour sessions and data collection}\label{sec:4-2}

Participants completed a paper pre-tour survey labeled with a pseudonymous ID while a researcher explained the procedure and helped install the app. They experienced the tour alone or in groups. The protocol asked participants to visit at least three of the seven stops. Stop selection reflected participants' interests, available time, weather and physical constraints, and researcher recommendations. Participants experienced different combinations of stops; recorded visits are summarized in Appendix~\ref{app:b2}.

A researcher accompanied participants, took written notes, and audio-recorded the session. After each stop, the researcher invited participants to share their thoughts or feelings, followed by stop-specific questions \citep{boren2000thinking}. At Names in the Sky, for example, participants were asked how the narrative and names displayed overhead made them feel about the mission. After early analysis identified spatial dissonance, researchers invited some subsequent visitors to describe the cemetery surroundings before opening the AR layer, probing how their initial readings related to the experience. Section~\ref{sec:5-1} reports one participant's reflection on this prompt.

At the final stop, participants reflected on the stops they had experienced and then completed the post-tour survey. Tour sessions lasted 20--50 minutes, excluding surveys. The Ohlone Youth Cultural Camp was intended for youth ages 12--18. Nine Ohlone participants recruited through the youth cultural camp---seven youth and two adults---also joined a collective debrief after their sessions, discussing the experience and possible future extensions. The debrief was audio-recorded and transcribed.

\subsection{Measures and survey analysis}\label{sec:4-3}

We piloted the survey with eight university community members and campus visitors, revising unclear protocol and phrasing as needed. The pre-tour survey recorded community affiliations, prior relationships with the site and its histories, prior AR use, touring-group size, demographics, and initial descriptions of the site. The full survey instrument is provided in the supplementary materials.

Six statements used a five-point response scale from 1 (strongly disagree) to 5 (strongly agree) before and after the tour (Figure~\ref{fig:pre-post}). Five were designed around the messages Ohlone collaborators wanted the tour to communicate. For example, ``This is Ohlone land'' informed ``I could describe specific places on this campus that are significant to Ohlone history.'' These items elicited self-assessed capabilities, not independently scored demonstrations of understanding. Each was accompanied by an open prompt requesting a concrete example. The sixth item asked whether Ohlone history was among the first things that came to mind when considering the site and had no accompanying open prompt.

Four post-tour statements addressed seeing the site differently, perceived Ohlone care and input, the importance of physical presence, and encountering a new perspective (Figure~\ref{fig:post-experience}). Two open prompts asked what remained with participants and what surprised them or prompted further curiosity. A later-added physical-markers item was excluded from quantitative analysis because of partial coverage. On forms containing it, this item preceded open reflections and may have prompted comments about physical acknowledgment.

For the 57 complete cases, we report item-level means and paired pre-to-post changes by public (Appendix~\ref{app:survey}). Participants with overlapping memberships appear once in each applicable public and once in pooled summaries. We also report descriptive comparisons by prior visits to the site. All quantitative results are descriptive.

\subsection{Qualitative corpus and analysis}\label{sec:4-4}

The qualitative corpus comprised 280 idea-level extracts from survey responses, recorded tour conversations and debriefs, and observation notes. Each extract was a response or passage treated as a unit for coding. All substantive survey open responses were included, encompassing the paired short-answer items and post-tour reflections; blank responses, refusals, and other non-substantive entries were excluded. Counts drawn from these responses refer to unique respondents whose coded answers contained the relevant idea or wording.

We conducted collaborative thematic analysis informed by Braun and Clarke's emphasis on researcher subjectivity and recursive interpretation \citep{braun2019reflecting, braun2021onesize}. Five researchers each familiarized themselves with and independently coded all 280 extracts in a shared spreadsheet, revisiting and refining their codes over successive passes. Across data collection, coding, thematic mapping, and manuscript drafting, team members participated in approximately ten interpretation sessions. We treated differences among researchers' readings as resources for interpretation, using these discussions to reconsider emerging interpretations and develop the analysis.

Following independent coding, we used Claude 5.0 Fable to propose consolidations of overlapping code labels, a technique with precedents in LLM-assisted code merging and aggregation \citep{gao2024collabcoder, chen2026computational}. One researcher reviewed every proposed consolidation, and original labels were retained alongside consolidated labels. The team then worked collectively with an interactive thematic map linking consolidated labels to original codes and supporting extracts. We compared similarities and differences across publics, revised relationships between codes and candidate themes, and developed analytic memos. We selected themes for reporting based on their relevance to the research questions, evidentiary depth, and attention to divergent cases. Appendix~\ref{app:coding}, Table~\ref{tab:code-development} illustrates code consolidation and the development of interpretations through this collective work.

\subsection{Ethics and positionality}\label{sec:4-5}

An institutional review board approved the study's consent and compensation procedures. Consent was recorded on the pre-tour survey, and declining ended participation. Mission visitors and university participants received \$20 gift cards, and powwow participants received \$20 cash. Ohlone camp leaders and youth received \$50 gift cards, consistent with prior compensation for community members' cultural knowledge and expertise.

Survey and recording files used pseudonymous IDs. Quotations use Ohlone-N, Native-N, Visitor-N, University-N, Ohlone/University-N, or University/Visitor-N, where N is a participant number. Composite labels preserve overlapping public memberships and remain consistent across quotations. Quotations from Ohlone youth in the collective debrief that could not be attributed individually use Ohlone-Youth.

\citet{kuhn2024indigenous} emphasize relationships, trust, and accountability in research with Tribal communities. The author team includes an Ohlone community member and non-Native researchers and practitioners. Existing relationships with Ohlone community members informed the study's evaluative priorities and interpretation of the tour's representational choices. The Ohlone co-author reviewed interpretations concerning Ohlone participants and cultural representations, while the team attended to negotiated and unfavorable reception across publics.

We treat the five messages as priorities articulated through co-design with contributing Ohlone partners, without claiming that they represent every Ohlone person \citep{noe2024generalized}. Where interpretation concerned memorial practice, ceremony, or language, we drew on the Ohlone co-author's expertise and ongoing communications with community members, while recognizing the limits of that knowledge. The non-Native authors do not speak for the community.

\section{Findings}\label{sec:5}

\begin{figure}[htbp]
\centering
\includegraphics[width=0.8\linewidth]{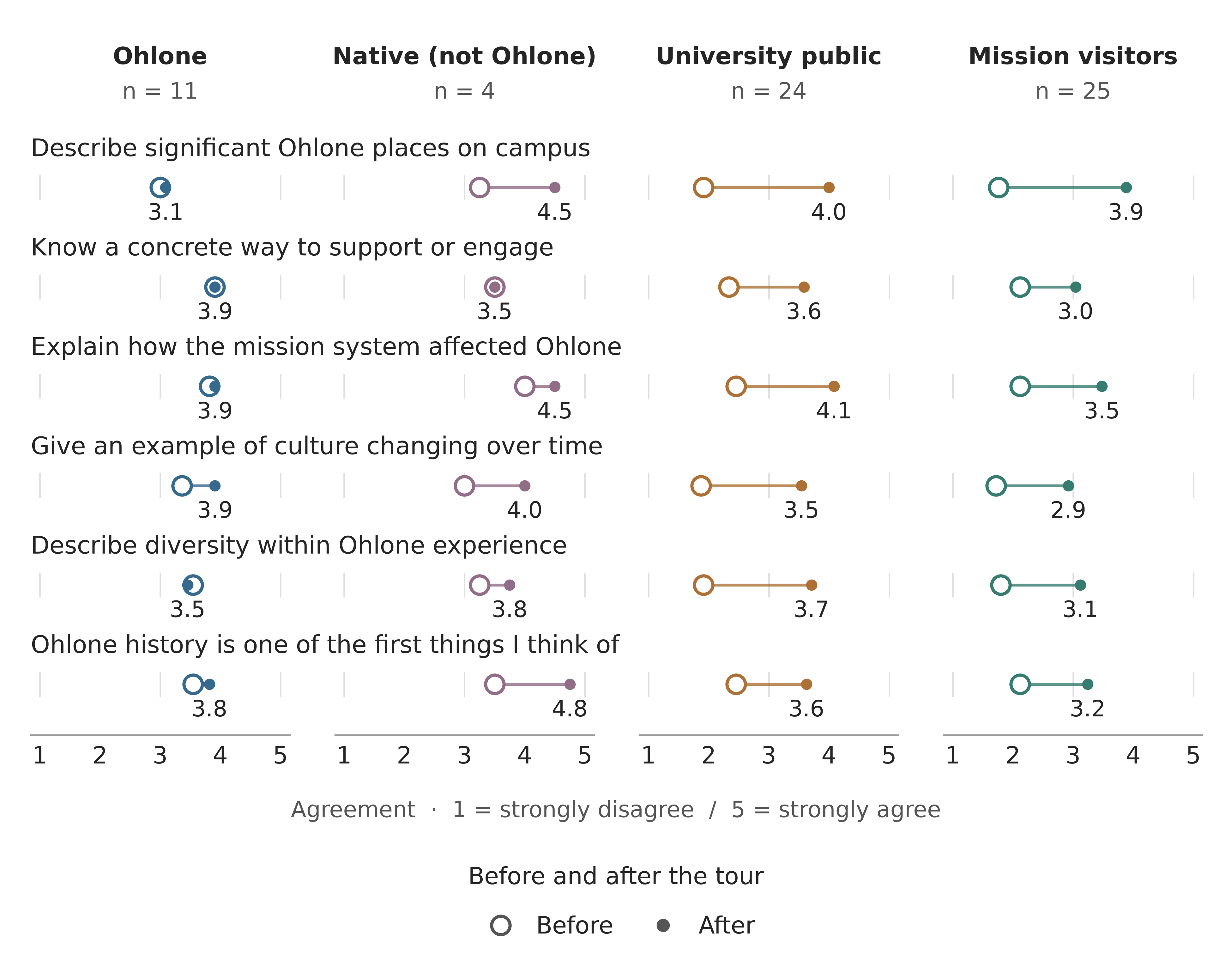}
\caption{Mean agreement before and after the tour, by group. Larger open circles show pre-tour means; smaller filled circles show post-tour means. Labels give post-tour means to one decimal place. The four Native participants who are not Ohlone are reported separately alongside the three publics. The analysis includes 57 unique respondents; public memberships overlap.}
\label{fig:pre-post}
\Description{Four columns compare mean agreement before and after the tour for six self-report items. Columns represent Ohlone participants (11), Native participants who are not Ohlone (4), the university public (24), and Mission visitors (25). Items concern describing significant Ohlone places, support or engagement, mission-system effects, cultural change, diversity within Ohlone experience, and Ohlone history coming to mind. Each column uses the same agreement scale from 1, strongly disagree, to 5, strongly agree. Larger hollow circles mark before means and smaller filled circles mark after means, with connecting lines and numeric after labels. All six means increase for the university and Mission-visitor groups. Five increase for the Native group, with support or engagement unchanged. Four increase for the Ohlone group; support or engagement is unchanged, and diversity decreases slightly from about 3.55 to 3.45. The two exact ties retain a visible outer ring. The 57 unique respondents contribute overlapping public memberships.}
\end{figure}

The publics began with different ratings, which generally increased after the tour. The six paired items comprised five self-assessments of capabilities related to the community-defined messages and one item about whether Ohlone history was among the first things that came to mind when considering the site. Across these items, Ohlone community members began between 3.0 and 3.9 and ended between 3.1 and 3.9; the university public began between 1.9 and 2.5 and ended between 3.5 and 4.1; Mission visitors began between 1.7 and 2.1 and ended between 2.9 and 3.9; the four Native participants began between 3.0 and 4.0 and ended between 3.5 and 4.8 (Figure~\ref{fig:pre-post}; detailed paired results in Table~\ref{tab:paired-items}, Appendix~\ref{app:a2}). Ohlone participants began with higher ratings, leaving less room for increases; smaller changes alone therefore do not indicate weaker reception. The four Native participants who are not Ohlone appear as a separate column throughout. With four people we read their responses as description rather than comparison, and we note where they diverge from Ohlone participants, since merging the two would have attributed that divergence to the Ohlone public. All group means increased or remained stable except Ohlone participants' ratings for describing diversity within Ohlone experience, which decreased slightly. Among the three publics, the university public showed the largest mean increase on five of six items, while Mission visitors had the lowest post-tour means on five of six. We report three findings, each pairing relevant survey items with what participants said and did at the tour stops: how relation to place shaped the encounter, how the messages traveled through what participants already knew and whom they met, and what participants wanted to change after the tour.

\subsection{Spatial dissonance: The tension between the visible campus and its AR layer}\label{sec:5-1}

\begin{figure}[htbp]
\centering
\includegraphics[width=\linewidth]{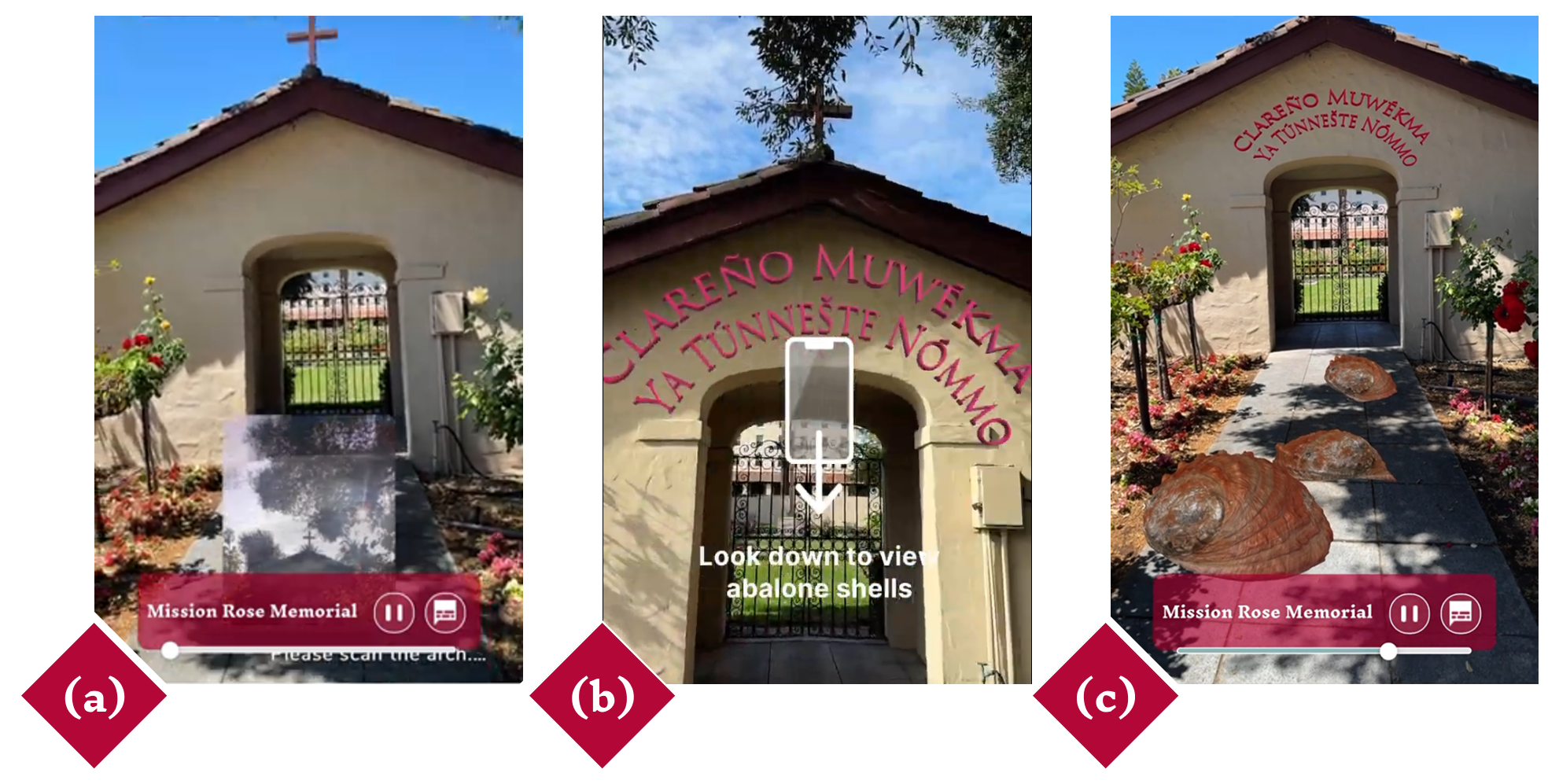}
\caption{The Cemetery Stop reveal: (a) the physical rose garden and gate without AR; (b) community-authored Chochenyo naming (Clareño Muwékma Ya Túnnešte Nómmo, meaning ``Where the Clareño Indians are Buried'') on the arch and the instruction to look down; (c) virtual abalone shells on the path underfoot. The app label ``Mission Rose Memorial'' refers to the same experience as this paper's ``Cemetery Stop''. The video figure in the supplementary materials demonstrates the Cemetery Stop’s AR content and narration.}
\label{fig:cemetery}
\Description{Three panels show the garden entrance through the app. Panel a shows the physical gate and an arch without the digital name or shells, with app controls overlaid near the bottom. Panel b places the community-authored Chochenyo name in red lettering above the arch and displays a phone-and-arrow symbol with an instruction to look down to view abalone shells. Panel c shows virtual abalone shells across the paved path leading to the named arch. The app's playback bar reads Mission Rose Memorial.}
\end{figure}

Participants across all three publics described how being at the site made the messages more meaningful. At the Ohlone summer-camp debrief, a youth explained: \emph{``the difference between a classroom, learning about the Holocaust and going to a museum in Auschwitz is the fact that hey, these people are literally underneath your feet. {[}\ldots{]} Like you\textquotesingle re standing on the ground where it happened''} (Ohlone-Youth). Six Ohlone, ten university, and six Mission visitor participants named the site itself as what made a message real. The survey item ``being physically present made the messages more meaningful'' also received high ratings across groups (means 3.8--5.0; Figure~\ref{fig:post-experience}).

Spatial dissonance arose from a felt contrast between the campus's visible presentation and the meanings participants encountered or brought to it. For university participants and some Mission visitors, the AR layer challenged everyday or Catholic readings of the campus. For Ohlone participants, it could affirm a familiar history while drawing attention to how little the physical campus acknowledged it.

University participants described reinterpreting familiar campus places through shock, sadness, and reflection. Everyday fixtures initially read as decorative or functional acquired second identities. The contrast was especially apparent at the Mission rose garden (Figure~\ref{fig:cemetery}): \emph{``the Rose Garden is a cemetery but I never thought of it like that before''} (University-11). One student momentarily forgot their familiar location outside the science building while learning about Ohlone history and culture, then recognized where they were again (University-18). Other accounts cast everyday campus practices in a more troubling light: sunbathing in the garden could feel disrespectful to people who knew its history (University/Visitor-2), while a student reflected uneasily on their upcoming performance of Catholic songs in the mission (University-3).

The survey also asked participants to name a place on campus connected to Ohlone history. University participants' pre-tour answers included the mission, while some left this prompt blank; after the tour, answers often named the cemetery or garden, sometimes combining both terms. On the corresponding rating item, agreement that one could describe places significant to Ohlone history rose from 1.9 to 4.0 for the university public. Agreement that Ohlone history was one of the first things that came to mind rose from 2.5 to 3.6 (Figure~\ref{fig:pre-post}). Among the three publics, university participants also gave the highest ratings for seeing the place differently and encountering an unfamiliar perspective (Figure~\ref{fig:post-experience}). Discomfort accompanied recognition of Ohlone histories in places participants used every day.

Most Mission visitors were on campus for the first time, encountering the site and its history together. Their ratings for describing places significant to Ohlone history rose from 1.8 to 3.9, similar to the university public, and they named the cemetery as it was introduced. Visitors also expressed unease: Visitor-13 described looking at the church while hearing that it was \emph{``still on top of bodies,''} calling this \emph{``a very depressing direction.''} Agreement that Ohlone history came first to mind rose from 2.1 to 3.2, the lowest post-tour level of any group. Repeat visitors rated seeing the place differently higher than first-time visitors did (4.2 against 3.5).

Visitors could also notice a contrast between their first impressions of a place and the history revealed by the tour. After early analysis identified spatial dissonance in university participants' responses, researchers asked some subsequent visitors to describe the Cemetery Stop before opening the AR layer. Asked afterward whether the prompt had affected the experience, one visitor explained: \emph{``since you asked me, then I had to really think about my environment vs walking passively through it''} (Visitor-14). This account suggests that describing the surroundings can make the contrast with the AR layer more salient.

Most Ohlone participants were youth, and their familiarity with the present-day campus varied. Some knew it well through previous camps, events, powwows, or visits with family members studying at the university; others were less familiar with it. They encountered representations of their own community's history on ancestral land, and their responses included kinship and grief alongside criticism of how little the campus acknowledged that history. Ohlone participants named the same sites before and after the tour, and their ratings moved little from already high baselines: place knowledge from 3.0 to 3.1, first-to-mind from 3.5 to 3.8. A community adult described the contrast between the garden's appearance and its ancestral significance: \emph{``They just see a bunch of beautiful roses, but it\textquotesingle s actually associated with a lot of death and tragedy for our people. So, we look at the beautiful roses today, but what it actually means is there\textquotesingle s thousands of our ancestors underground''} (Ohlone/University-1). The speaker described the app as a way of making that history visible. The account combined ancestral connection and grief with concern about what the physical campus acknowledged.

Limited acknowledgment also drew criticism beyond the tour stops. In a campus parking garage, a pavement outline and small plaque mark part of a former adobe dormitory that housed Native families at Mission Santa Clara \citep[p.~47]{ohwg2020report}. Although the garage was not a tour stop, Ohlone youth who passed by it during the tour later discussed its marker in the debrief: \emph{``something this significant is just tucked away in the corner''} (Ohlone-Youth). Ohlone participants evaluated representations of their own community's history while bringing different ages, experiences, and levels of familiarity to that evaluation.

Participants' accounts also brought the question of ``why here?'' into focus. During the debrief, one Ohlone youth asked why the fountain stop's general account of seasonal practices belonged there: \emph{``Realistically, you could write that about any location. Like, why specifically did you choose the fountain?''} (Ohlone-Youth). At the cemetery, participants could connect the history directly to the ground beneath their feet. Across the three publics, that connection could bring unfamiliar histories into view, change the meaning of everyday surroundings, or affirm ancestral relationships. Being there mattered through the relationships participants recognized between Ohlone lives and that particular place.

\subsection{Familiar stories and recognition of Ohlone life today}\label{sec:5-2}

Participants interpreted the tour through familiar histories and religious or commemorative conventions. These interpretive repertoires shaped how they understood Ohlone accounts, while encounters with contemporary people and practices connected those accounts to Ohlone life today.

Participants drew on familiar histories to interpret the mission system. Ratings of the ability to explain its effects on Ohlone communities rose from 2.5 to 4.1 for the university public and from 2.1 to 3.5 for Mission visitors. Most Mission visitors reported at least moderate familiarity with mission history, making their low initial mean for explaining its effects on Ohlone people particularly notable. Participants related missionization to slavery: it \emph{``sounded like how slavery worked''} (University-10). Visitor-24 compared the erasure of Ohlone history with attempts under Hitler to \emph{``overwrite another culture\textquotesingle s history.''} An Ohlone youth drew on the difference between learning about the Holocaust in a classroom and visiting Auschwitz to explain what physical presence added: \emph{``you\textquotesingle re standing on the ground where it happened''} (Ohlone-Youth). Some brought expectations about colonial harm: \emph{``it\textquotesingle s not like I\textquotesingle m surprised, because it\textquotesingle s not like our country really cared about Native Americans''} (Visitor-14).

At Names in the Sky, familiar commemorative and religious conventions supported different readings. A mission staff member connected the memorial with annual commemorations of martyrs: \emph{``you know the impact based on the number, but when you start to hear stories about folks, it personalizes people a little bit''} (University/Visitor-2). A student objected that the names overhead were \emph{``consistent with Christian ideology when the whole point of the audio was the forced involvement by the Missions''} (University-5). A companion offered another reading: \emph{``their memory hangs over the current space''} (University-6). University-16 judged the hard-to-read names through a familiar convention of respect: \emph{``everything\textquotesingle s displayed like on a plaque, it\textquotesingle s very clear, orderly, and like that\textquotesingle s how we show respect.''} Another visitor interpreted the memorial through faith: \emph{``It makes me feel that their spirits are still here, around and that I\textquotesingle m glad they were baptized. So that means they\textquotesingle re in heaven''} (Visitor-16). Familiar conventions could support connection with the people being remembered, criticism of the representation, or reinterpretation through the visitor's own beliefs.

Ohlone participants connected the tour to living cultural practices and relationships. Ohlone-8 emphasized both continuity and change: \emph{``Culture\textquotesingle s alive and well {[}\ldots{]} It\textquotesingle s very modified, but as is everything.''} The same speaker rejected the frozen-at-the-mission picture outright: \emph{``We\textquotesingle re not a static people {[}\ldots{]} Why cook in a basket when I got an iron pot, you know?''} In her survey responses, she asked for more about today's community and people, including ceremony, regalia, and social activity. Other participants foregrounded personal connections and contemporary lives. Ohlone-9 described what stayed with them as \emph{``How I connect with my roots.''} Asked what surprised them or what they wanted to learn more about, Ohlone-6 named \emph{``Students attending and how the college supports them personally.''}

For university participants and Mission visitors, contemporary people helped convey the message ``we are still here.'' Eight university and five Mission visitor participants named a person, a voice, or the students as what stayed with them. University-13 contrasted the encounter with textbook accounts: \emph{``It\textquotesingle s good to see actual children and now to meet actual Native Americans at SCU. I just always hear about them in textbooks and stuff.''} A visitor described \emph{``a blend of historical and what people are doing now, you know. These people still exist''} (Visitor-12). Students also recognized peers. University-9 valued hearing another student's perspective and \emph{``first-hand experience with events and doing stuff actively on-campus that\textquotesingle s really unheard of because the school doesn\textquotesingle t really promote that stuff.''} A visitor emphasized intergenerational continuity: \emph{``It\textquotesingle s so good to know that things are being passed on''} (Visitor-17), while another valued seeing the narrators interacting and playing a game (Visitor-15).

Digital objects also connected accounts to family relationships and ongoing practice. University-15 recalled a narrator's grandfather's staves and a weaver's grandmother's basket; University-17 described the basket as \emph{``still a technology''} for Ohlone people. These accounts located cultural practices in relationships between living people and earlier generations.

These connections varied across stops, including within the same participant's account. Visitor-16, who interpreted Names in the Sky through Christian beliefs about baptism and heaven, welcomed the Powwow as a celebration by Ohlone people on campus: \emph{``Everyone\textquotesingle s in the present''} and \emph{``bringing together the past and the present.''} Other accounts left connections unresolved. At the same Basket Technology Stop that prompted University-15 to recall a family relationship, Visitor-14 described the experience as \emph{``the least related to people''} and questioned how it conveyed their presence. Visitor-12 wanted to know \emph{``what happened to those Ohlone generations between the mission times and modern times.''}

University participants' and Mission visitors' ratings for giving an example of cultural change and describing differences in Ohlone experience increased by 1.2--1.8 points (Figure~\ref{fig:pre-post}). Their accounts show how recognition of a living Ohlone community could coexist with familiar religious interpretations and questions about the connections between historical narratives and community life today.

\subsection{``Up there for real'': Calls for physical acknowledgment}\label{sec:5-3}

Calls for physical acknowledgment appeared across the three publics. Eight university participants, five Mission visitors, and two Ohlone participants called for signs, plaques, memorials, or names that would make Ohlone history visible on campus. Visitor-14 wanted a marker that \emph{``clearly states, the Ohlone people, not just Native Americans''}, while University-6 called for signage identifying the rose garden as a burial site. At the Cemetery Stop, the digital lettering above the garden gate (Figure~\ref{fig:cemetery}b) prompted a request for a physical version: \emph{``It made my heart expand, and I think it should be up there for real, not digital''} (Visitor-15).

For Ohlone participants, acknowledgment connected commemoration with what the community carries forward. \emph{``It\textquotesingle s good to have representation and for names to be seen,''} said Ohlone-6. A community adult emphasized \emph{``what we\textquotesingle re keeping alive and passing on to the next generation''} (Ohlone/University-1). These concerns aligned with an existing effort: the Ohlone History Working Group, which included Ohlone leaders, had recommended plaques and interpretive panels at campus burial sites in 2020 \citep{ohwg2020report}.

Requests for further acknowledgment coexisted with appreciation of existing efforts. University-18 expressed pride in the Scholarship Stop's account of institutional support, but contrasted it with the cemetery: \emph{``you\textquotesingle re hiding the history, you\textquotesingle re not really talking about it. If I didn\textquotesingle t take this AR tour, I wouldn\textquotesingle t know''}. Visitor-14 likewise valued the university doing something to address the past while asking for a plaque. For Visitor-17, the absence of familiar burial markers made \emph{``the sacredness of that space''} unclear; at Names in the Sky, the same visitor asked, \emph{``Why are we being quiet?''}. These accounts evaluated both the steps the university had taken and what remained publicly unacknowledged.

Survey responses showed positive perceptions of community involvement alongside more varied ratings of knowing how to engage. Forty-three of 57 respondents (75\%) agreed or strongly agreed that the tour was \emph{``created with genuine care and input from the Ohlone community''}. Agreement that one could identify a concrete way to support or engage with the community remained at 3.9 for Ohlone participants and 3.5 for the four Native participants who were not Ohlone. It rose from 2.3 to 3.6 for the university public and from 2.1 to 3.0 for Mission visitors (Figure~\ref{fig:pre-post}). Even after the tour, 16 of 25 Mission visitors remained neutral or disagreed on this item.

Some participants named possible next steps in their open responses: supporting campus clubs (University-4), attending the Powwow (University-5), explaining the significance of campus places to others (University-14), bringing middle-school students to visit (Visitor-22), and donating toward a physical marker (Visitor-24). These were proposed ways to engage; the study did not follow whether participants pursued them.

Physical markers and AR could also be valued together. Visitor-14 requested a plaque while describing AR's superimposed elements as making the experience more immersive, with the qualification that \emph{``you have to make sure people actually have the app''}. Participants could appreciate what the digital encounter offered while asking for acknowledgment that people could encounter without opening the app.

\section{Discussion}\label{sec:6}

The same AR layer was received differently across the site\textquotesingle s publics. Participants reconsidered familiar places, encountered histories connected to their own lives, and interpreted the messages through existing cultural references. Across these differences, appreciation of the tour accompanied requests for acknowledgment beyond the app. These findings connect three design tasks: staging an encounter with the place, making representational choices understandable, and developing public support for community-proposed changes.

\subsection{Stage spatial dissonance for emotional impact}\label{sec:6-1}

Here, spatial dissonance refers to the felt tension between a site's visible presentation and meanings people bring to or encounter there.

\citet[pp.~20--21]{tunbridge1996dissonant} locate heritage dissonance in the selective production of heritage, through which some pasts are privileged and others excluded. Levine's \emph{San Francisco--Baghdad} project brings a distant ``there'' into a local ``here'' by mapping bombing sites in Baghdad onto San Francisco \citep[p.~147]{levine2014common}. Her \emph{Berlin Wall} example also makes an absent structure visible at its original location \citep[pp.~149--150]{levine2014common}. Our study focuses on reinterpreting the ``here'': tensions between the garden's visible presentation, its significance as an Ohlone burial place, and its public acknowledgment. Community-authored AR could challenge familiar readings or affirm ancestral connections while exposing inadequate acknowledgment. Our contribution is to examine comparatively how these tensions are experienced by people with different relationships to the same place.

Learning or recognition alone need not produce that tension. The design opportunity is to support reflection on the felt mismatch. This connects to two design resources discussed in the related work: defamiliarization, making familiar things available for renewed attention \citep{bell2005making}, and ambiguity, which invites people to interpret a situation through more than one frame \citep{gaver2003ambiguity}. The AR layer can draw attention to a tension already present in the landscape: an attractive garden belies its significance as a burial place.

The observation-prompt episode reported in the findings suggests a way to pair surprise with reflection. Visitor-14 said that the invitation to describe the garden prompted closer attention to the surroundings before opening the AR layer. A precedent comes from classroom demonstrations, where asking students to predict an outcome before seeing it improved conceptual understanding \citep{crouch2004classroom}. Inviting visitors to articulate an initial reading of a place could similarly give them something explicit to reconsider after encountering the Ohlone account.

We propose staging the encounter as observation, reflection, reveal, and return. Before opening the layer, the app could invite visitors to describe the place and what it means to them. Visitors could reflect privately or offer a brief description, then start the layer themselves. At its end, a proposed prompt could ask them to lower the phone, look again, and consider: ``What do you notice about this place now?'' This sequence could support reflection whether the encounter challenges a familiar reading or exposes inadequate acknowledgment of a meaning already held.

The return step gives participants time to consider how the account relates to the surroundings they still see. \citet{benford2012uncomfortable} similarly place emotionally challenging moments within a wider sequence that includes reflection afterward. Where visitors choose to share their readings, evaluation could ask whether anything felt inconsistent, how they responded to the encounter, and whether their understanding of the place or assessment of its public acknowledgment changed.

\subsection{``We are still here'': Connect histories to community life today}\label{sec:6-2}

\citet{harrison1996replacing} locate the meaning of place in communities' practices and cultural understandings. Our findings show how audiences encountered these relationships through community-authored AR. The peer recognition and family-object anecdotes reported in the findings connected campus history to fellow students and practices carried between generations. For Ohlone participants, contemporary community life was already familiar and something they wanted more fully represented. These differences suggest designing accounts that support both recognition of existing relationships and discovery of relationships visitors had not previously considered.

Read through \citeauthor{vizenor2008survivance}'s \citep{vizenor2008survivance} account of survivance, these encounters draw attention to Native people actively carrying forward and changing cultural life. The design opportunity is to give those activities a substantive role in heritage narratives. Building on the Ohlone voices already present throughout the tour, accounts could describe how people learn a practice, adapt it, share it with others, and connect it to their lives today. The Ohlone speaker's comparison of a cooking basket with an iron pot makes this point directly: cultural life includes adopting new tools and changing familiar practices. Designing for ``we are still here'' means representing a contemporary, evolving community whose members decide what to retain and what to change.

Those activities also help explain why an account belongs at a particular site. \citet[p.~7]{schroeder2023location} describe AR narration that connects speakers' experiences to buildings visible around the visitor. Our findings suggest examining how audiences understand the community relationships conveyed through such connections. At other heritage sites, narration could explain how a location figures in people's lives: as a gathering place used today, the setting of an ancestral account, or a place chosen for commemoration. Community partners could determine which relationships deserve attention and how to represent them.

Making these relationships explicit still leaves room for interpretation. \citet{gaver2003ambiguity} value ambiguity for inviting people to develop their own readings; \citet{hall1980encoding} likewise shows why reception cannot be prescribed by a message's producer. The same visitor's differing responses to Names in the Sky and the Powwow Stop illustrate how familiar beliefs remained active across encounters. Evaluation could ask participants how they understand the community's relationship to a site and what that relationship means to them. Discussing these responses with community partners could inform revisions where an important connection was overlooked, while preserving space for different judgments about the account.

\subsection{Prototype public acknowledgment and support advocacy}\label{sec:6-3}

Participants' calls for physical acknowledgment suggest how reception can help communities turn future imaginaries into proposals for changing a place \citep{lewis2023future}. At the Cemetery Stop, the digital Chochenyo-language cemetery name above the garden gate (Figure~\ref{fig:cemetery}b) prompted a visitor to ask that it be ``up there for real, not digital'' (Visitor-15). This builds on AR creators' accounts of prototyping realities they wanted to build together \citep[p.~4]{silva2022understanding}.

At contested heritage sites, designers could work with community partners to make alternative futures available for consideration on the ground where they would take shape. Proposed names, memorials, or artworks could be compared in relation to existing buildings, routes, and monuments. Community review could address whose experiences each proposal represents and what it leaves unresolved. \citet{cardenasgasca2025lessons} likewise emphasize continuing community influence over portrayals of their histories and possible futures, alongside collaborative AR authoring. At stages chosen by community partners, reception among other publics could reveal how proposals are understood and where their meanings require clarification.

Preliminary discussions among Ohlone community members have considered a statue of Ohlone dancers, in line with the \citeauthor{ohwg2020report}'s \citep{ohwg2020report} recommendation for a statue or other artwork recognizing Ohlone people and culture. Such a statue would add an Ohlone representation to a campus that already features dozens of statues of Catholic saints. Questions remain about the number and genders of dancers, whether they represent particular people or general figures, and who would create the work. AR could help community members compare compositions, scale, and placement through successive rounds of discussion. The prototype would make choices tangible while remaining open to revision through community deliberation.

Connecting these representations to advocacy would give visitors ways to act on the futures they encounter. \citet{schroeder2023location} describe a virtual plaque whose design they believe informed a later physical installation. Designers at other sites could make this relationship an explicit design question: what change is proposed, who can bring it about, and how can interested publics contribute? Depending on the initiative, an AR experience or associated outreach could invite petition signatures, participation in consultations, or financial support. In this project, the Scholarship Stop could link to donations supporting the Ohlone and Muwekma Ohlone Scholarship Fund, while the Powwow Stop could invite visitors to the next annual powwow. Evaluation could then follow how reception informs revisions and commitments, and whether institutional responses change what is acknowledged on the ground.

\subsection{Limitations}\label{sec:6-4}

Our study captures interpretations formed during a single encounter. We cannot establish whether spatial dissonance prompted lasting reflection, was reconciled with familiar narratives, or faded from attention. Later visits and retellings could reveal whether participants retained the specificity of Ohlone accounts or absorbed them into existing cultural frames.

We examined one configuration of community-authored narration, spatially anchored representations, and place. Participants' accounts illuminate how these elements were experienced together, but do not establish what the visual overlays contributed beyond encountering the same stories on site through another medium. The proposed observation--reflection--reveal--return sequence likewise remains a direction for design and evaluation.

Finally, calls for permanent acknowledgment leave open what support would look like around particular designs, locations, funding commitments, or institutional responsibilities. We did not follow proposals through community deliberation and institutional decision-making, so cannot establish how the tour might contribute to realizing them. Yet Dear Visitor offers a hopeful precedent: its creators believe their virtual plaque informed a subsequent physical installation \citep{schroeder2023location}, suggesting that site-based AR can help advance community-advocated change.

\section{Conclusion}\label{sec:7}

Comparing reception of the Thámien Ohlone AR Tour across multiple publics showed how relationships to place and cultural frames informed encounters with community-authored accounts. Spatial dissonance ran in different directions: AR could unsettle familiar campus readings or affirm Ohlone histories while making their limited physical acknowledgment more troubling. Contemporary people and their stories made Ohlone continuance tangible, while appreciation of the tour accompanied calls for institutional action. These findings suggest using AR to stage reflection, clarify representational choices, and prototype forms of public acknowledgment under community direction. A rose garden can also be a cemetery. By making that coexistence visible, community-authored AR can invite people to reconsider where they stand---and offer communities another way to ask what acknowledgment of that ground should look like.

\FloatBarrier
\bibliographystyle{ACM-Reference-Format}
\bibliography{references}

\clearpage
\appendix
\numberwithin{table}{section}
\numberwithin{figure}{section}
\renewcommand{\thetable}{\thesection\arabic{table}}
\renewcommand{\thefigure}{\thesection\arabic{figure}}
\section{Detailed survey results}\label{app:survey}

\subsection{Analytic sample and reporting}\label{app:a1}

The primary survey analysis includes 57 participants who completed the six paired items and four common post-tour items. We report the three overlapping publics separately: Ohlone community members (n = 11), university community members (n = 24), and Mission visitors (n = 25). Four Native participants who did not identify as Ohlone community members are reported separately alongside these publics. Their responses are not included in the Ohlone group or treated as a fourth public.

Fifty-three participants account for the 60 memberships across the three publics: three belong to both the Ohlone and university publics, and four to both the university and Mission-visitor publics. Each appears once in every applicable public and once in pooled summaries. The four Native participants who are not Ohlone bring the unique-participant total to 57.

All items use a five-point scale from 1 (strongly disagree) to 5 (strongly agree). The tables use abbreviated item labels; the full survey instrument is provided in the supplementary materials. These study-specific items are reported individually and do not constitute a validated composite scale. Results are descriptive; we performed no imputation and report no significance tests. A later-added item about physical markers was excluded from analysis because of partial coverage. On forms containing it, the item preceded open-ended reflections and may have prompted comments about physical acknowledgment.

\subsection{Paired survey items}\label{app:a2}

\begin{table}[htbp]
\caption{Pre-tour and post-tour agreement with the six paired survey items. Each cell gives the pre-tour mean $\rightarrow$ post-tour mean, followed by the mean within-person change in parentheses. Changes were calculated from unrounded paired responses and then rounded to two decimal places, so they can differ from subtraction of the displayed means. These are self-assessments, not independently scored demonstrations of knowledge.}
\label{tab:paired-items}
\centering
\small
\setlength{\tabcolsep}{3pt}
\renewcommand{\arraystretch}{1.18}
\begin{tabularx}{\linewidth}{@{}>{\raggedright\arraybackslash}p{.31\linewidth}*{4}{>{\centering\arraybackslash}X}@{}}
\toprule
\textbf{Item} & \textbf{Ohlone}\newline $(n=11)$ & \textbf{Native, not Ohlone}\newline $(n=4)$ & \textbf{University}\newline $(n=24)$ & \textbf{Mission visitors}\newline $(n=25)$ \\
\midrule
Describe significant Ohlone places on campus & \shortstack{$3.00\rightarrow3.09$\\$(+0.09)$} & \shortstack{$3.25\rightarrow4.50$\\$(+1.25)$} & \shortstack{$1.92\rightarrow4.00$\\$(+2.08)$} & \shortstack{$1.76\rightarrow3.88$\\$(+2.12)$} \\
\addlinespace
Identify a concrete way to support or engage with Ohlone culture and community & \shortstack{$3.91\rightarrow3.91$\\$(0.00)$} & \shortstack{$3.50\rightarrow3.50$\\$(0.00)$} & \shortstack{$2.33\rightarrow3.58$\\$(+1.25)$} & \shortstack{$2.12\rightarrow3.04$\\$(+0.92)$} \\
\addlinespace
Explain how the mission system affected Ohlone communities & \shortstack{$3.82\rightarrow3.91$\\$(+0.09)$} & \shortstack{$4.00\rightarrow4.50$\\$(+0.50)$} & \shortstack{$2.46\rightarrow4.08$\\$(+1.63)$} & \shortstack{$2.12\rightarrow3.48$\\$(+1.36)$} \\
\addlinespace
Give an example of Ohlone culture changing over time & \shortstack{$3.36\rightarrow3.91$\\$(+0.55)$} & \shortstack{$3.00\rightarrow4.00$\\$(+1.00)$} & \shortstack{$1.88\rightarrow3.54$\\$(+1.67)$} & \shortstack{$1.72\rightarrow2.92$\\$(+1.20)$} \\
\addlinespace
Describe diversity within Ohlone cultural experience & \shortstack{$3.55\rightarrow3.45$\\$(-0.09)$} & \shortstack{$3.25\rightarrow3.75$\\$(+0.50)$} & \shortstack{$1.92\rightarrow3.71$\\$(+1.79)$} & \shortstack{$1.80\rightarrow3.12$\\$(+1.32)$} \\
\addlinespace
Ohlone history is among the first things that come to mind about this place & \shortstack{$3.55\rightarrow3.82$\\$(+0.27)$} & \shortstack{$3.50\rightarrow4.75$\\$(+1.25)$} & \shortstack{$2.46\rightarrow3.63$\\$(+1.17)$} & \shortstack{$2.12\rightarrow3.24$\\$(+1.12)$} \\
\bottomrule
\end{tabularx}
\end{table}

\subsection{Common post-tour items}\label{app:a3}

\begin{figure}[htbp]
\centering
\includegraphics[width=\linewidth]{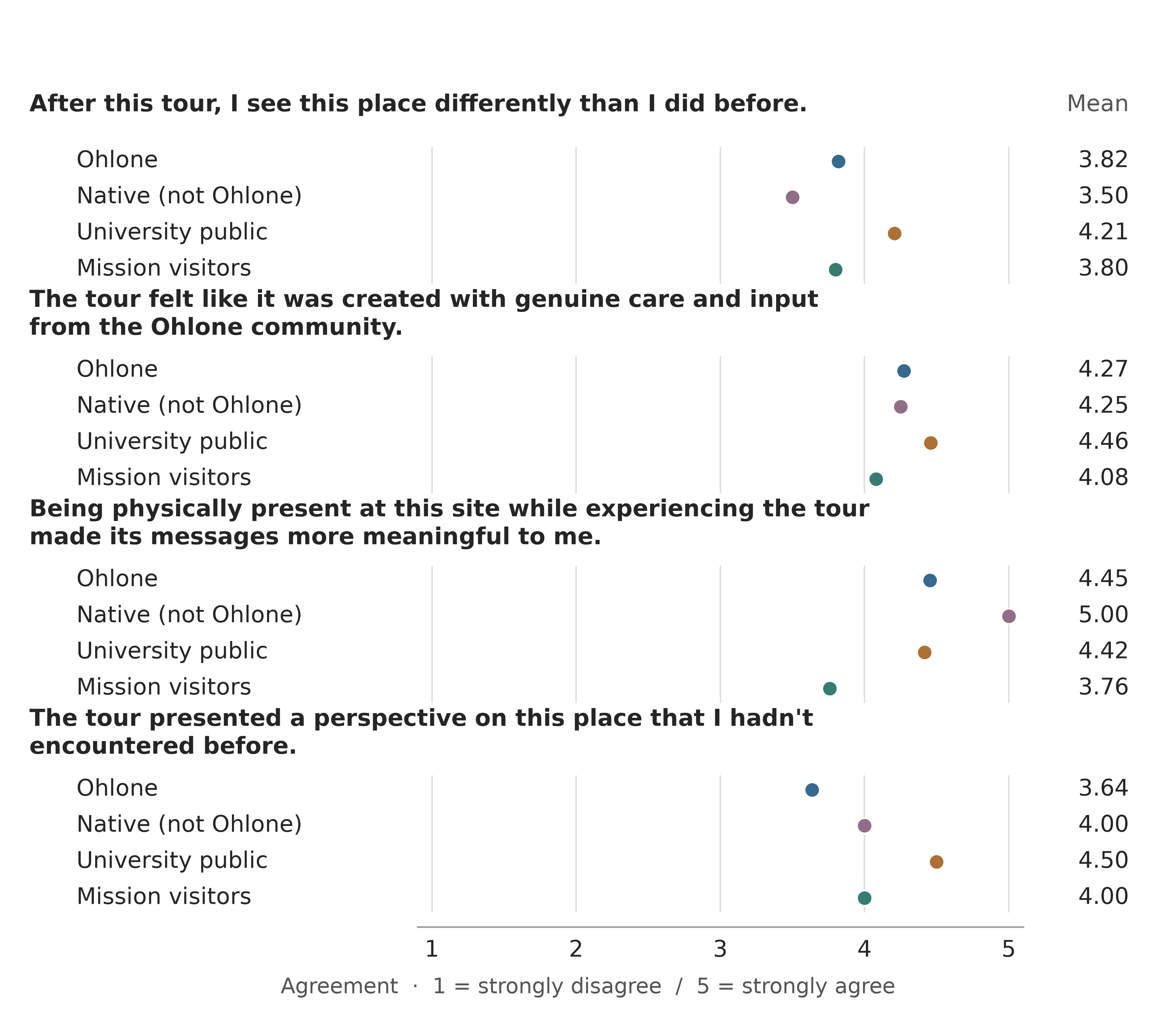}
\caption{Mean agreement with the four common post-tour items, by group: Ohlone ($n=11$), Native participants who are not Ohlone ($n=4$), the university public ($n=24$), and Mission visitors ($n=25$). Labels give means to two decimal places; points use unrounded values. The response scale is 1 (strongly disagree) to 5 (strongly agree). Native participants who are not Ohlone are reported separately alongside the three publics. The analysis includes 57 unique respondents; public memberships overlap as described in A.1.}
\label{fig:post-experience}
\Description{A dot plot displays four full survey statements. Within each block the group order is Ohlone, Native participants who are not Ohlone, the university public, and Mission visitors. Agreement runs from 1, strongly disagree, to 5, strongly agree. Means appear in a right-hand column. For seeing the place differently after the tour, means in this group order are 3.82, 3.50, 4.21, and 3.80. For the tour feeling created with genuine care and input from the Ohlone community, they are 4.27, 4.25, 4.46, and 4.08. For physical presence making the messages more meaningful, they are 4.45, 5.00, 4.42, and 3.76. For encountering a new perspective on the place, they are 3.64, 4.00, 4.50, and 4.00. Points use unrounded means. There are 57 unique respondents and overlapping public memberships.}
\end{figure}

\section{Survey coverage and recorded stop visits}\label{app:coverage}

\subsection{Survey coverage}\label{app:b1}

Table~\ref{tab:survey-coverage} summarizes participant groups and survey completeness. Qualitative material from the four participants with incomplete primary surveys remained eligible for analysis when available.

\begin{table}[htbp]
\caption{Participant groups and survey coverage.}
\label{tab:survey-coverage}
\centering
\small
\setlength{\tabcolsep}{4pt}
\renewcommand{\arraystretch}{1.15}
\begin{tabularx}{\linewidth}{@{}>{\raggedright\arraybackslash}p{.38\linewidth}*{3}{>{\centering\arraybackslash}X}@{}}
\toprule
\textbf{Participant group} & \textbf{Full survey sample} & \textbf{Complete primary survey} & \textbf{Incomplete primary survey} \\
\midrule
Ohlone community members & 11 & 11 & 0 \\
\addlinespace
Native participants who were not Ohlone & 4 & 4 & 0 \\
\addlinespace
University community members & 25 & 24 & 1 \\
\addlinespace
Mission visitors & 28 & 25 & 3 \\
\midrule
\textbf{Unique participants} & \textbf{61} & \textbf{57} & \textbf{4} \\
\bottomrule
\end{tabularx}
\par\smallskip
\begin{minipage}{\linewidth}
\footnotesize\textit{Note. Three participants belonged to both the Ohlone and university publics, and four to both the university and Mission-visitor publics; group rows therefore should not be summed. Complete primary surveys contain responses to the six paired items and four common post-tour items. Three participants provided no post-tour survey and one provided a partial post-tour survey.}
\end{minipage}
\end{table}

Before the tour, six of 11 Ohlone respondents, four of 24 university respondents, and two of 25 Mission visitors reported being very or extremely familiar with Ohlone history and presence in the area. All respondents in the primary survey sample answered this item.

\subsection{Recorded stop visits}\label{app:b2}

All 57 participants in the primary survey sample had a nonempty post-tour stop checklist. Table~\ref{tab:stop-visits} combines these checklists with one researcher-confirmed correction: a Native participant who was not Ohlone visited the Scholarship stop but omitted it from the checklist. A researcher who accompanied the participant confirmed the visit. These records do not establish that every part of a stop\textquotesingle s content was successfully experienced; one Basket Technology visit was explicitly marked as broken.

\begin{table}[htbp]
\caption{Recorded visits to each tour stop among participants in the primary survey sample. Columns give counts of participants, except the final row, which gives the median number of stops and range. Participants could visit multiple stops, and public memberships overlap. The pooled column counts each person once.}
\label{tab:stop-visits}
\centering
\small
\setlength{\tabcolsep}{3pt}
\renewcommand{\arraystretch}{1.15}
\begin{tabularx}{\linewidth}{@{}>{\raggedright\arraybackslash}p{.25\linewidth}*{5}{>{\centering\arraybackslash}X}@{}}
\toprule
\textbf{Stop} & \textbf{Ohlone}\newline $(n=11)$ & \textbf{Native, not Ohlone}\newline $(n=4)$ & \textbf{University}\newline $(n=24)$ & \textbf{Mission visitors}\newline $(n=25)$ & \textbf{All participants}\newline $(n=57)$ \\
\midrule
Thámien Village & 10 & 3 & 13 & 11 & 31 \\
Names in the Sky & 6 & 3 & 20 & 25 & 48 \\
Powwow & 6 & 3 & 12 & 22 & 39 \\
Cemetery & 6 & 2 & 21 & 22 & 44 \\
Basket Technology & 3 & 1 & 12 & 22 & 33 \\
Staves & 2 & 0 & 7 & 8 & 14 \\
Scholarship & 9 & 1 & 14 & 9 & 28 \\
\midrule
Median number of stops (range) & 3 (2--7) & 3 (3--4) & 4 (3--7) & 4 (2--7) & 4 (2--7) \\
\bottomrule
\end{tabularx}
\end{table}

Although the protocol requested at least three stops, recorded visits ranged from two to seven. Stop selection reflected participants' interests, available time, weather and physical constraints, and researcher recommendations; exposure was not assigned experimentally. Differences in responses therefore cannot be attributed to public membership independently of the content participants encountered.

\section{Code consolidation and theme development}\label{app:coding}

\begin{table}[htbp]
\caption{Illustrative examples of code consolidation, collective interpretation, and connections to the findings.}
\label{tab:code-development}
\centering
\small
\setlength{\tabcolsep}{4pt}
\renewcommand{\arraystretch}{1.15}
\begin{tabularx}{\linewidth}{@{}>{\raggedright\arraybackslash}p{.32\linewidth}>{\raggedright\arraybackslash}p{.23\linewidth}>{\raggedright\arraybackslash}X@{}}
\toprule
\textbf{Examples of original code labels} & \textbf{Consolidated code} & \textbf{Development through collective interpretation} \\
\midrule
``physical visibility wanted''; ``wants permanent physical marker''; ``need for burial signage'' & ``mark it for real: signage, plaques, a memorial'' & Considered alongside ``an app can't stand in for a plaque'' in developing the \textbf{Permanence: Mark It for Real} memo. Comparing appreciation of the tour with requests for lasting markers informed Section~\ref{sec:5-3}'s account of physical acknowledgment. \\
\addlinespace
``visualizing precolonization makes it `realer'''; ``ar visualization makes past concrete''; ``ar aiding historical visualization'' & ``seeing it here makes it real'' & Considered alongside ``standing where it happened'' and ``a physical place makes it count'' in developing the \textbf{Being There Makes It Real} memo. Read together with comparisons of participants' prior relationships to place, this informed Section~\ref{sec:5-1}'s account of how being there made histories meaningful and brought the campus's limited acknowledgment into view. \\
\addlinespace
``living culture not just past''; ``appreciation of living culture''; ``recognition of contemporary native existence'' & ``we're still here'' & Developed in the \textbf{We Are Still Here (Continuance Carried by People)} memo by comparing Ohlone participants' accounts of cultural continuity with university participants' recognition of contemporary Ohlone peers and Mission visitors' varied readings. These comparisons informed Section~\ref{sec:5-2}'s account of how contemporary voices and practices conveyed living culture, alongside cases where that connection remained unresolved. \\
\bottomrule
\end{tabularx}
\end{table}

One researcher reviewed every proposed consolidation. The examples illustrate selected connections developed through collective thematic mapping; individual codes could contribute to more than one candidate theme.

\end{document}